\documentclass[aps,twocolumn,nopacs,preprintnumbers]{revtex4}

\usepackage{graphicx, fancyhdr, amsfonts, amsmath, amssymb, hyperref, glossaries, siunitx}
\usepackage{pdfpages}
\hypersetup{
    colorlinks=true,
    linkcolor=blue,
    filecolor=magenta,      
    urlcolor=cyan,
    pdftitle={Overleaf Example},
    pdfpagemode=FullScreen,
    }
\usepackage[T1]{fontenc}
\usepackage{xr}
\usepackage{color,soul}
\newcommand{\subtxt}[2]{\ensuremath{#1_\mathrm{#2}}}

\newcommand{\sige}[0]{\gls{si}\gls{ge}}

\newacronym{dqd}{DQD}{double quantum dot}
\newacronym{qd}{QD}{quantum dot}
\newacronym{qw}{QW}{quantum well}
\newacronym{soi}{SOI}{spin--orbit interaction}
\newacronym{hh}{HH}{heavy-hole}
\newacronym{lh}{LH}{light-hole}
\newacronym[\glslongpluralkey={two dimensional hole gases}]{2dhg}{2DHG}{two dimensional hole gas}
\newacronym{haadf}{HAADF}{high-angle annular dark-field}
\newacronym{pcb}{PCB}{Printed Circuit Board}
\newacronym{vna}{VNA}{Vector Network Analyzer}
\newacronym{hemt}{HEMT}{High Electron Mobility Transistor}

\newacronym{hs}{HS}{heterostructure}
\newacronym{sshh}{SSHH}{superconductor/semiconductor hybrid heterostructure}
\newacronym{sige}{SiGe}{silicon-germanium}
\newacronym{ge}{Ge}{germanium}
\newacronym{si}{Si}{silicon}
\newacronym{pt}{Pt}{platinum}
\newacronym{al}{Al}{aluminum}
\newacronym{cu}{Cu}{copper}

\newacronym{hf}{HF}{hydrofluoric acid}
\newacronym{rt}{RT}{Room Temperature}
\newacronym{bcs}{BCS}{Bardeen–Cooper–Schrieffer}

\newacronym{sem}{SEM}{scanning electron microscopy}
\newacronym{tem}{TEM}{transmission electron microscopy}
\newacronym{cvd}{CVD}{chemical vapor deposition}
\newacronym{ald}{ALD}{atomic layer depostion}
\newacronym{rie}{RIE}{reactive ion etching}
\newacronym{sdh}{SdH}{Shubnikov–de Haas oscillations}
\newacronym{dhg}{2DHG}{two-dimensional hole gas}
\newacronym{edx}{EDX}{energy dispersive X-Ray}
\newacronym{icp}{ICP}{inductively coupled plasma}
\newacronym{tlf}{TLF}{two-level fluctuators}
\newacronym{flp}{FLP}{Fermi-level pinning}
\newacronym{vb}{VB}{valence band}
\newacronym{fl}{E$_F$}{Fermi-level}
\newacronym{alo}{Al$_2$O$_3$}{aluminum oxide}
\newacronym{sin}{SiN$_x$}{silicon nitride}
\newacronym{eb}{E$_b$}{thermal-activation energy}
\newacronym{np}{n$_p$}{percolation density}
\newacronym{cnl}{CNL}{charge neutrality level}
\newacronym{mos}{MOS}{metal-oxide semiconductor}
\newacronym{str}{s}{strained}
\newacronym{cpw}{CPW}{coplanar waveguide}
\newacronym{tl}{TL}{transmission line}
\newacronym{tls}{TLS}{two-level systems}
\newacronym{qp}{QP}{quasiparticle}
\newacronym{qps}{QPs}{quasiparticles}
\newacronym{dr}{DR}{dilution refrigerator}
\newacronym{mbe}{MBE}{molecular-beam epitaxy}
\newacronym{squid}{SQUID}{Superconducting Quantum Interference Device}

\newacronym{qi}{\subtxt{Q}{i}}{internal quality factor}

\DeclareSIUnit\torr{Torr}
\DeclareSIUnit\evolt{eV}
\DeclareSIUnit\sccm{sccm}
\DeclareSIUnit{\square}{\ensuremath{\Box}} 

\DeclareSIUnit\dbm{dBm}
\DeclareSIUnit\degreekelvin{K}

\begin{document}

\title{High-quality and field resilient microwave resonators on Ge/SiGe quantum well heterostructures}

\author{Luigi Ruggiero\textsuperscript{1}}
\author{Carlo Ciaccia\textsuperscript{1}}
\author{Pauline Drexler\textsuperscript{2}}
\author{Vera Jo Weibel\textsuperscript{1}}
\author{Christian Olsen\textsuperscript{1}}
\author{Christian Schönenberger\textsuperscript{3, 4}}
\author{Dominique Bougeard\textsuperscript{2}}
\author{Andrea Hofmann\textsuperscript{1, 3, 5}}

\begin{abstract} 
\begin{center}
\textsuperscript{1}  University of Basel, Klingelbergstrasse 82, 4056 Basel, Switzerland \\
\textsuperscript{2} Fakultät für Physik, Universität Regensburg, Universitätsstraße 31, 93053 Regensburg, Deutschland \\
\textsuperscript{3} Swiss Nanoscience Institute, Klingelbergstrasse 82, 4056 Basel, Switzerland\\
\textsuperscript{4} YQuantum, Parkstrasse 1, 5234 Villigen, Switzerland\\
\textsuperscript{5} Corresponding author, email: andrea.hofmann@unibas.ch\\
\vspace{1em}
\today
\end{center}

Superconducting resonators integrated with \gls{ge} \glspl{qw} offer a promising platform for hybrid quantum devices. Yet, in the most common heterostructure architectures, they have so far been limited by sizable photon losses. Here, we report the fabrication and characterization of microwave resonators patterned in the \gls{al} thin film of an in-situ grown superconductor/semiconductor hybrid \gls{hs}. The semiconductor part of this hybrid \gls{hs} is grown on a commercial \gls{ge} substrate. We consistently achieve internal quality factors $\subtxt{Q}{i}>1000$, surpassing previous results on Ge \gls{qw} heterostructures grown using the concept of a virtual \gls{ge} substrate on \gls{si} substrates. We reach $\subtxt{Q}{i}\approx49000$ at single-photon occupation and a plateau of $\subtxt{Q}{i}\approx20000$ at sub-one photon, an order of magnitude larger than any previously reported value of resonators on \gls{ge} QW structures at low power. We further characterize the thin \gls{al} film forming the resonator, extracting its kinetic inductance and superconducting gap, and studying its magnetic field dependence. Notably, the resonance remains well-defined up to in-plane magnetic fields of \SI{850}{\milli\tesla}. With a hysteresis emerging in the out-of-plane magnetic field dependence, for both the resonance frequency and the quality factor, indicating an interesting interplay between vortex- and quasiparticle loss mechanisms.
\end{abstract}

\maketitle

\section{Introduction}
Strained \gls{ge} \glspl{qw} are a promising platform for both spin qubits \cite{hendrickx_fast_2020, hendrickx_four-qubit_2021, jirovec_singlet-triplet_2021} and hybrid quantum devices \cite{aggarwal_enhancement_2021, vigneau_germanium_2019, hendrickx_ballistic_2019}. Integration of these devices with microwave resonators \cite{de_palma_strong_2024, janik_strong_2025, sagi_gate_2024, kiyooka_gatemon_2025, nigro_demonstration_2024} is attractive as it enables fast, high-fidelity readout and offers a route toward long-range qubit coupling via cavity photons \cite{mi_strong_2017, stockklauser_strong_2017, dijkema_cavity-mediated_2025}. However, combining a compact heterostructure architecture with high-quality resonators has been challenging so far, mainly due to the lossy substrate originating from specific growth techniques. The strained \gls{ge} \gls{qw} is built by growing a thin layer of \gls{ge} in-between two strain-relaxed \gls{ge}-rich \sige\ layers. Starting from a commercial \gls{si} substrate, the unstrained \sige\ layer is obtained via a strain-relaxing buffer \cite{fitzgerald_dislocations_1991}, either by gradually increasing the \gls{ge} content, or by first growing a pure \gls{ge} layer and subsequent decreasing of the \gls{ge} content. This latter approach has recently gained increased attention due to the reported high mobilities and successful qubit operations performed on such heterostructures \cite{sammak_shallow_2019, hendrickx_four-qubit_2021}. However, it contains many dislocations at the \gls{si}-\gls{ge} interface, due to the large lattice mismatch of the two materials. The microwave losses introduced by these defects limit the reported \subtxt{Q}{i} to about \num{1000} \cite{nigro_demonstration_2024}. On the other hand, larger \subtxt{Q}{i} values of a few thousand have been achieved \cite{sagi_gate_2024, de_palma_strong_2024,janik_strong_2025, kiyooka_gatemon_2025} by the former approach, at the cost of thick buffer layers and long growth times. Additionally, most studies on microwave resonators on Ge have so far been limited to low fields, even though analyses on field resilience \cite{janik_strong_2025} are crucial for qubit applications.

Here, we present an approach delivering high \subtxt{Q}{i} and a large in-plane magnetic field resilience for resonators which are structured into a \gls{mbe} grown superconductor/semiconductor hybrid \gls{hs}. Different from many existing \gls{ge}/\gls{sige}\ \glspl{hs} \cite{hendrickx_fast_2020, hendrickx_four-qubit_2021, jirovec_singlet-triplet_2021,aggarwal_enhancement_2021, vigneau_germanium_2019, hendrickx_ballistic_2019,de_palma_strong_2024, janik_strong_2025, sagi_gate_2024, kiyooka_gatemon_2025, nigro_demonstration_2024}, our hybrid \gls{hs} is grown on a commercial \gls{ge} substrate with appropriate grading to a constant \sige\ composition \cite{stehouwer_germanium_2023}. This allows us to avoid the dislocation density-rich \gls{si}-\gls{ge} interface, which provides loss-channels for the resonator \cite{de_palma_strong_2024}, and still keep the \gls{hs} stack compact. The hybrid \gls{hs} is furthermore characterized by an in-situ grown thin superconducting \gls{al} film, brought onto the semiconductor without breaking the vacuum. To characterize the material platform and its suitability for microwave technology, we systematically compare eight resonators of varying lengths and investigate the dependence of $Q_i$ and resonance frequency on applied magnetic fields.

 
\section{Experimental}
\label{sec:results}

\subsection{Hybrid heterostructure and setup}
\label{sec:substrate}
\begin{figure*}[ht!]
        \centering
        \includegraphics[width = 1.0\textwidth]{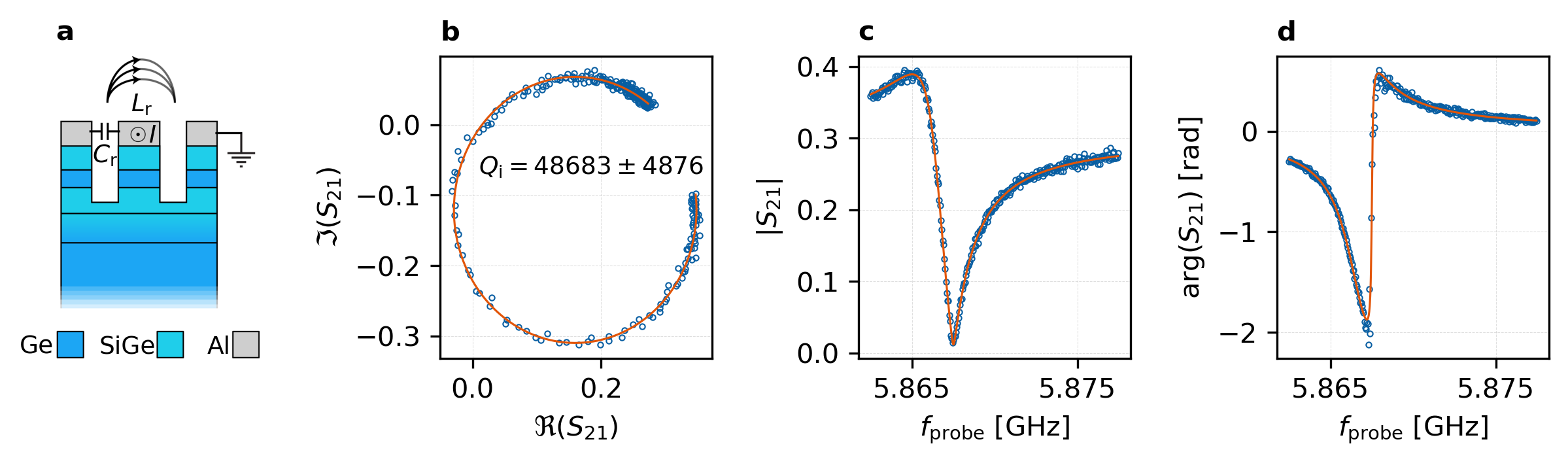}
        \caption{\textbf{High-Q resonator on a \gls{ge} \gls{qw}}. (a) Cross-sectional schematic (not to scale) of a microwave resonator patterned from an \gls{al} film deposited on a \gls{ge}/\sige\ quantum well, which is epitaxially grown on a \gls{ge} substrate. (b - d) Fit (orange solid line) of the I/Q, transmission amplitude and phase response, respectively. The extracted quality factor exceeds \num{4.8e4} at an average circulating photon number of $<\subtxt{n}{p}> \simeq 1.0$.}  
        \label{fig:best}
\end{figure*}

The \gls{mbe}-grown hybrid superconductor/semiconductor \gls{hs} used here features an \SI{8}{\nano\meter} thin \gls{al} film directly grown onto a \gls{ge}/\sige\ \gls{hs} with a \gls{ge} \gls{qw}. The whole stack, including the \SI{475}{\micro\meter} thick commercial \gls{ge} substrate, is shown in \autoref{fig:best}(a). Detailed information on the growth can be found in the Supplementary Information. The \gls{tl} and the resonators are carved out of the \gls{al} film by standard Transene-D wet-etching of the \gls{al} followed by \gls{rie} of the \gls{hs}. The \gls{ge}/\sige\ \gls{qw} remains intact below the resonator. 

We measure three samples: Sample A with eight $\lambda/4$ resonators hanging to a \gls{tl}, and samples B and C with a single $\lambda/4$ resonators hanging to a \gls{tl}. In all the samples, the \gls{tl} is much longer than the resonators and designed to be \SI{50}{\ohm}-matched at both ends (see Supplementary Information) in order to avoid interference of standing waves. The experiment is designed such that we can separately estimate the internal quality factor \subtxt{Q}{i} and the coupling quality factor \subtxt{Q}{c}, from which we extract the loaded quality from $1/\subtxt{Q}{l} = 1/\subtxt{Q}{i} + 1/\subtxt{Q}{c}$. Unless specified otherwise, the experiments are carried out in a \gls{dr} with a base temperature of \SI{10}{\milli\kelvin}, using a Rohde\&Schwarz ZNB8 vector network analyzer. The input line is attenuated inside the \gls{dr} allowing for low powers down to below one average circulating photon in the resonator. We measure the transmitted signal through the \gls{tl} \subtxt{S}{21}, which travels through a superconducting coaxial cable up to \SI{4}{\kelvin}, where it is amplified by a Low Noise Factory LNF\_LNC1\_12A \gls{hemt} amplifier, the signal is amplified again at room temperature with an additional amplifier Miteq AMF-3F-01000400-08-10P.

In a first set of experiments, sample A is cooled down in the \gls{dr} with a passive \gls{al} magnetic field shield. The ground plane is connected through several \gls{al} bonds to the ground of the cryostat. Additional bonds bridging across each resonator and the \gls{tl} aim to further reduce potential inhomogeneities of the ground plane. The \gls{pcb} is embedded in a \gls{cu} box in order to minimize the impact of standing waves. 

With this setup, we achieve an internal quality factor $\subtxt{Q}{i}=\SI{48683\pm4876}{}$ at low input power $\subtxt{P}{in}$ corresponding to an average number $<\subtxt{n}{p}> = (2 \subtxt{P}{in}\subtxt{Q}{l}^{2})(\hbar\pi\subtxt{f}{r}\subtxt{Q}{c}) \simeq 1$ photon circulating in the resonator \cite{weber_single_2011}. The internal \subtxt{Q}{i} and external quality factor \subtxt{Q}{c}, as well as the resonator resonance frequency, are extracted from a circular fit \cite{probst_efficient_2015} of the resonator response in the complex plane, as shown in \autoref{fig:best}(b - d). The resonance frequency, \subtxt{f}{r}, has been determined by measuring the complex transmission as a function of frequency, where the absolute value shows a dip and the phase response shows a characteristic jump at resonance, see \autoref{fig:best}(b) and (c), respectively. The extracted \subtxt{f}{r} amounts to \SI{5.86}{\giga\hertz} and the coupling quality factor is \subtxt{Q}{c} = $\SI{2531\pm21}{}$. Although here, the coupling quality factor limits the photon lifetime, the design can easily be adjusted to a higher \subtxt{Q}{c}. Meanwhile, the extracted \subtxt{Q}{i} corresponds to a photon lifetime of $\subtxt{Q}{i}/(2\pi \subtxt{f}{r}) \simeq$ \SI{2.6}{\micro\second}, which exceeds the coherence times of many reported qubits in Ge \cite{hendrickx_fast_2020,hendrickx_four-qubit_2021, jirovec_dynamics_2022}.

\subsection{Film characterization}
\label{subsec:best}
In contrast to other experiments performed on \gls{ge}/\sige\ \glspl{hs} where ex-situ deposited \gls{al} layers are used to induce a proximity effect \cite{valentini_parity-conserving_2024}, our resonators are directly fabricated from a hybrid \gls{hs} (see Supplementary Information). The \gls{mbe}-approach provides a high degree of control of the \gls{al} film thickness and allows to form an abrupt superconductor/semiconductor interface (see \gls{tem} in Supplementary Information), via crystalline growth in an ultra-high vacuum (UHV) environment. 

\begin{figure}[h!]
        \centering
        \includegraphics[width = 0.5\textwidth]{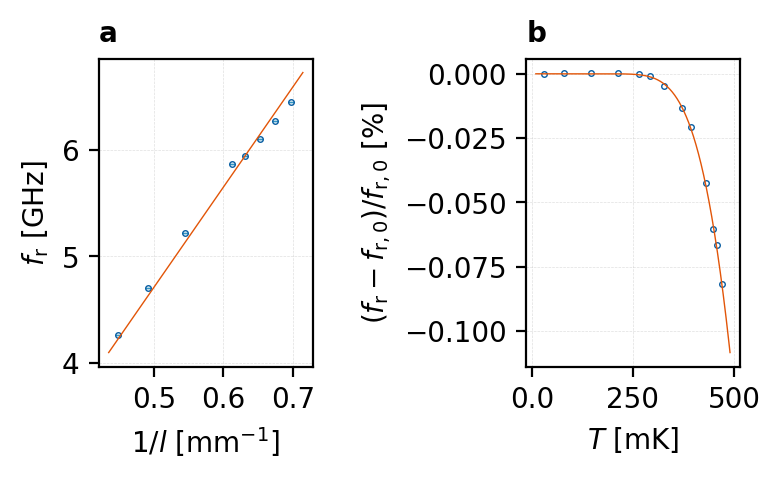}
        \caption{\textbf{Thin film properties extracted from resonator characteristics.} (a) The resonance frequency \subtxt{f}{r} is inversely proportional to the resonator length $l$. The data (blue dots) are fitted (solid line) to the equation $\subtxt{f}{r}(l)$ specified in the main text, yielding a kinetic inductance of \SI{36\pm2}{\pico\henry/\square} (b) The relative shift of the resonance frequency \subtxt{f}{r} (blue dots) measured at elevated temperature compared to $\subtxt{f}{r,0}:=\subtxt{f}{r,T=\SI{10}{\milli\kelvin}}$. The solid line is a fit to the \gls{bcs} theory (see main text), which  predicts a decreasing resonance frequency due to the lower density of Cooper pairs which increases the kinetic inductance.}
        \label{fig:Lkin_Temp}
\end{figure}

We first characterize the kinetic inductance \subtxt{L}{kin}, of the \gls{al} film. To do so, we measure the dependence of the resonance frequency on the resonator's length $l$, as shown in \autoref{fig:Lkin_Temp}(a). We fit our data to $\subtxt{f}{r}(l)=1/4l\sqrt{L_{\mathrm{r}} C_{\mathrm{r}}}$ valid for the employed quarter-wavelength geometry. Here, $L_{\mathrm{r}} = L_{\mathrm{geo}} + L_{\mathrm{kin}}$, $\subtxt{L}{geo}= (\mu_{\mathrm{0}}/4)(K(k'_{\mathrm{0}})/K(k_{\mathrm{0}})) = \SI{26.6}{\nano\henry/\meter}$ is the geometric inductance per unit length, $C_{\mathrm{r}} = 4\epsilon_{\mathrm{0}} \epsilon_{\mathrm{eff}} K(k'_{\mathrm{0}})/K(k_{\mathrm{0}})$ is the capacitance per unit length \cite{gevorgian_cad_1995}, and $\epsilon_{\mathrm{eff}}$ the material's dielectric constant. Using the dielectric constant and kinetic inductance as fitting parameters, we extract $\subtxt{\varepsilon}{eff} = \SI{10.01\pm0.01}{}$ and $\subtxt{L}{kin} = \SI{36\pm2}{\pico\henry/\square}$. With our \SI{8}{\nano\meter} \gls{al} film, we find that the kinetic inductance is four times larger than reported for a \SI{10}{\nano\meter} thick \gls{al} film grown ex-situ onto a \gls{qw} \gls{hs} in ultra-high vacuum environment \cite{valentini_parity-conserving_2024}, consistent with the fact that \subtxt{L}{kin} increases for thinner superconducting films.

To further characterize the \gls{al} film in terms of its superconducting properties, we analyze the resonance frequency as a function of temperature. It follows a \gls{bcs}-like behavior, where the Cooper pair density decreases at elevated temperature, increasing the kinetic inductance and reducing the resonance frequency. We fit $(\subtxt{f}{r} - \subtxt{f}{r,\SI{10}{\milli\kelvin}})/\subtxt{f}{r,\SI{10}{\milli\kelvin}} = -\alpha/2 \sqrt{\pi \Delta_{\mathrm{0}}/(2 \subtxt{k}{B} T)} \exp{(-\Delta_{\mathrm{0}}/(\subtxt{k}{B} T))}$ to the data shown in \autoref{fig:Lkin_Temp}(b) \cite{turneaure_surface_1991, grunhaupt_loss_2018}, where \subtxt{f}{r,\SI{10}{\milli\kelvin}} refers to the resonance frequency at \SI{10}{\milli\degreekelvin} and $\alpha = \subtxt{L}{kin,0}/(\subtxt{L}{kin,0} + \subtxt{L}{geo})$ is the kinetic inductance fraction at zero temperature. We extract a superconducting gap at zero temperature $\Delta_{\mathrm{0}} = \SI{298\pm4}{\micro\evolt}$, slightly higher than in a 10\,nm thick \gls{al} film grown ex-situ onto a QW \gls{hs} in UHV environment \cite{valentini_parity-conserving_2024}. Further, we extract the kinetic inductance fraction $\alpha=\SI{0.75\pm0.08}{}$ from the fit, in agreement with the value obtained from calculating $\subtxt{\alpha}{sim}=0.85$ from the inductance values \subtxt{L}{kin} and \subtxt{L}{geo}.

\begin{figure*}[ht!]
        \centering
        \includegraphics[width = 1.0\textwidth]{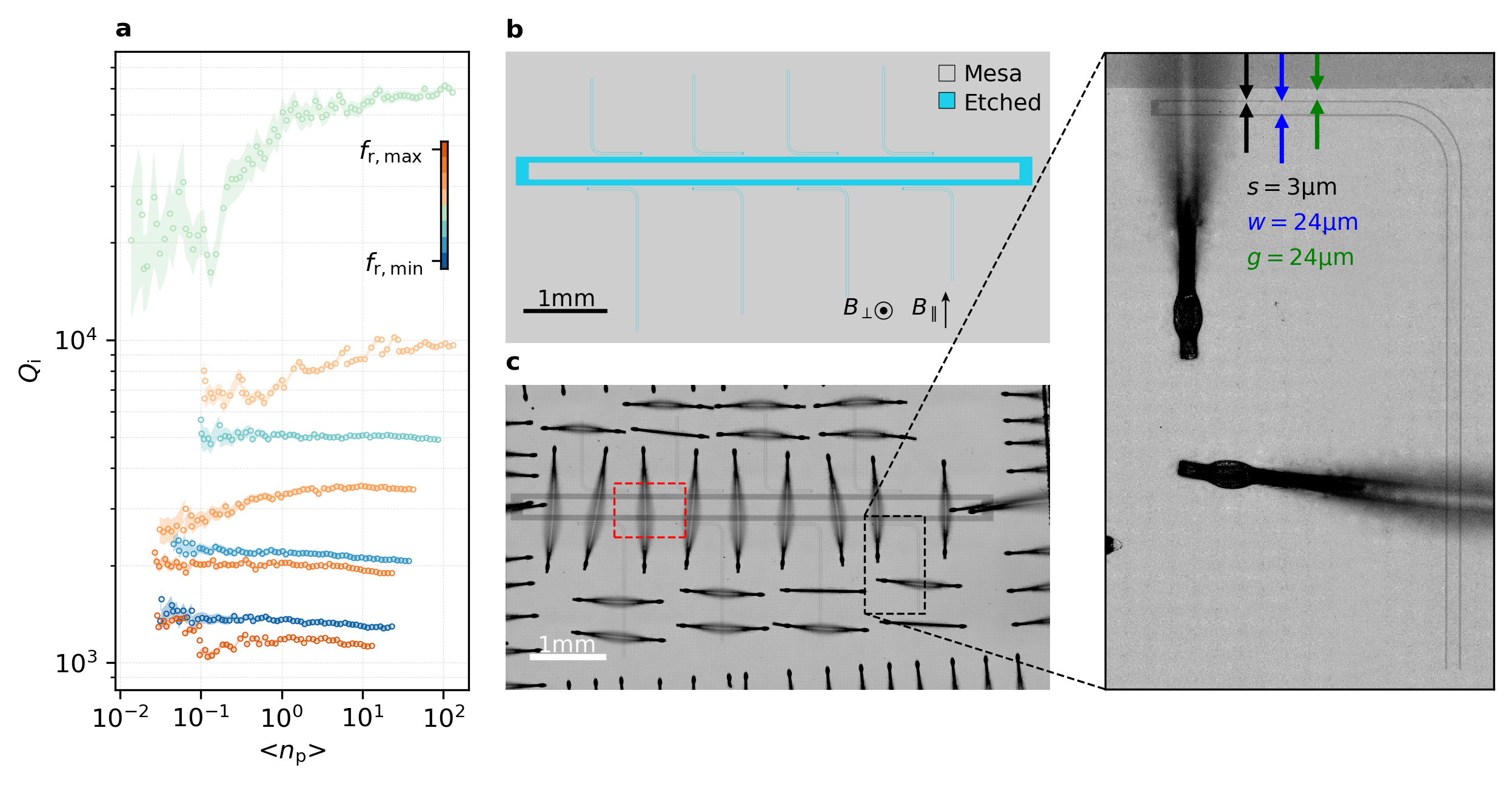}
        \caption{\textbf{Photon number dependence of eight $\lambda/4$ resonators connected to a single \gls{tl}}. (a) Quality factor as a function of average photon number for eight resonators of different frequencies ranging from about \SIrange{4}{7}{\giga\hertz}. (b) Design scheme with eight $\lambda/4$ resonators of different length $l$ hanging to a common \gls{tl}. (c) Top view of the sample on which the design in (b) is patterned. The impedance of the resonators at resonance frequency is $\subtxt{Z}{r} = \sqrt{\subtxt{L}{r}/{\subtxt{C}{r}}}$ $\simeq$ \SI{66}{\ohm} \cite{gevorgian_cad_1995}. The constant coupling length of $\SI{473}{\micro\meter}$ (not shown) and its relative variation to resonator length yields a variable coupling quality factor \subtxt{Q}{c} ranging from \SIrange{1200}{9000}{} in agreement with the simulations shown in SFig.2(b). It is also affected by grounding inhomogeneities, which we try to minimize by bonds connecting different areas of the ground plane. The area marked by red dashed line indicates a region of inhomogeneous etching.  (c.inset) Zoom-in on the resonator with the highest quality factor, with details on the feedline width $w=\SI{24}{\micro\meter}$, gap to ground $s=\SI{3}{\micro\meter}$, coupling $g =\SI{24}{\micro\meter}$, constant for all eight resonators.}
        \label{fig:Qi}
\end{figure*}

\subsection{Statistics on \subtxt{Q}{i}}
To gain insight into the variability of our resonators, we analyze and compare the performance of eight different resonators all patterned on the same sample A.
In \autoref{fig:Qi}(a) we plot \subtxt{Q}{i} for all eight resonators as a function of average photon number. A sketch of the design and an optical image of the bonded device are shown in \autoref{fig:Qi} (b) and (c), respectively.
The inset of \autoref{fig:Qi}(c) highlights the position of the best resonator relative to the whole sample described in \autoref{subsec:best}, showing all relevant dimensions. The chosen design yields impedance values slightly above \SI{50}{\ohm}, and we consistently measure quality factors exceeding \num{1000}. Thus, higher than what previously reported for resonators on \gls{ge}/\sige\ \glspl{hs} grown via \gls{ge} virtual substrates on commercial \gls{si} substrates\cite{nigro_demonstration_2024} and 
similar to those on \glspl{hs} grown directly on commercial \gls{si} \cite{valentini_parity-conserving_2024, sagi_gate_2024}. This corroborates previous suspicions \cite{nigro_demonstration_2024} that the defects at the interface between the \gls{si} substrate and the \gls{ge} virtual substrate are the main source of loss. 

The origin of the losses can be better understood by sweeping the input power circulating in the resonator. The \subtxt{Q}{i} of some of the resonators exhibits an increase with power before saturating at higher levels. This behavior is characteristic of resonator loss dominated by \gls{tls}, which cause dissipation at low photon numbers \cite{khalil_analysis_2012}. As the power increases, the \gls{tls} gradually saturate, reducing their contribution to loss and resulting in a plateau in \subtxt{Q}{i} at high power. The onset of this increase depends on factors such as the resonance frequency, the coupling strength to \gls{tls}, and the internal quality factor at low photon numbers. For most of our resonators, we observe this increase in \subtxt{Q}{i} already at relatively low powers. Notably, in the resonator with the highest observed \subtxt{Q}{i}, we also detect a plateau at low photon number, indicating a regime where unsaturated \gls{tls} dominate the loss \cite{khalil_analysis_2012}. In this case, \subtxt{Q}{i} plateaus at approximately \num{20000} at average photon numbers well below one.

For some of the resonators, we observe no pronounced dependence of \subtxt{Q}{i} on input power, suggesting that loss mechanisms other than \gls{tls} dominate \cite{nigro_demonstration_2024}. One possible source is an etching imperfection in the \gls{tl}, as indicated by a dashed red rectangle in \autoref{fig:Qi}(c), which may disrupt the homogeneity of ground currents or promote quasiparticle excitations. The relative significance of \gls{tls} loss---quantified by the difference in \subtxt{Q}{i} between low and high power or \subtxt{Q}{i, max} ---clearly correlates with the resonator's proximity to this defect (see Supplementary Information). Additionally, the placement of bond wires has been shown to influence resonator performance by altering the distribution of ground currents \cite{chen_fabrication_2014, zikiy_high-q_2023}. Nevertheless, all the resonators consistently display high internal quality factors above 1000.

\subsection{Impact of magnetic field}

For future qubit applications, information on the response of the resonators to magnetic fields will be important. We study the in-plane characteristics using sample B, which contains a resonator with \subtxt{f}{r}=\SI{3.76}{\giga\hertz} and \subtxt{Q}{i, B_\parallel=0}=\SI{1150}{} measured at base temperature, high photon number and zero applied field. Note that vortices in the magnet coil likely decrease the quality factor from its true zero-field value. The design of the transmission line is the same as used for sample A, and the coupling length of the same order. The in-plane field, defined as shown in \autoref{fig:Qi} (b)
, is produced by a vector magnet, which allows for a correction of a small sample tilt estimated to $\theta \simeq$ \SI{0.12}{\degree} by adjusting different axes of the vector magnet during the sweep.

In \autoref{fig:field}, we plot the shift in resonance frequency, $\delta\subtxt{f}{r} = \subtxt{f}{r}(B_\parallel=0) - \subtxt{f}{r}(B_\parallel>0)$ and the normalized quality factor $\subtxt{Q}{i}(B_\parallel>0) / \subtxt{Q}{i}(B_\parallel=0)$. The resonance frequency starts to shift significantly only at fields above \SI{100}{\milli\tesla}, due to lower density of Cooper pairs increasing the kinetic inductance. Regarding the quality factor, we choose to plot relative values, as the presence of vortices in the vector magnet reduces the zero-field \subtxt{Q}{i} compared to measurements performed in a magnet-free, shielded setup. The quality factor tends to decrease with increasing field, though its behavior is heavily affected by the fluctuations of the vector magnet and the vortices trapped in the superconducting film. Nevertheless, the quality factor has decreased by less than \SI{75}{\percent} at \SI{850}{\milli\tesla}. At this field, we can still perform a reliable fit to the dip in the transmission amplitude.
\begin{figure*}[ht!]
        \centering
        \includegraphics[width = 1.0\textwidth]{Figures_Sketches/Fig4.png}
        \caption{\textbf{Magnetic field dependence.}  (a) Shift of the resonance frequency and (b) normalized quality factor as a function of in plane magnetic field from zero to $\subtxt{B}{\parallel} =\SI{850}{\milli\tesla}$. (c) Shift of the resonance frequency and (d) normalized quality factor as a function of out-of plane magnetic field from zero to $\subtxt{B}{\perp} =\SI{50}{\micro\tesla}$. The arrows and the color (orange for increasing fields and blue for decreasing fields, respectively) indicate the direction of the sweep.}
        \label{fig:field}
\end{figure*}

The resonator’s behavior in out-of-plane magnetic fields is studied using sample C, which carries a resonator with \subtxt{f}{r}=\SI{3.45}{\giga\hertz} and \subtxt{Q}{i, B_\perp=0} $\simeq$ \SI{5000}{}. The results are shown in \autoref{fig:field}(c-d), showing a pronounced hysteresis. While the best possible quality factors are likely obtained by releasing vortices after each change of the magnetic field, such a scenario is not realistic for actual (qubit) device measurements. Hence, it is important to analyze the hystersis, which we believe is caused by an interplay between vortex dynamics in the superconducting film and quasiparticle excitation and trapping. The out-of plane field is supplied by a homemade superconducting coil mounted on the back of the \gls{pcb}, all the components are encapsulated in an \gls{al} magnetic field shield. Starting from zero field, the resonance frequency decreases gradually with increasing field. This is partly due to a narrowing of the effective width of the superconducting region in the center line, and partly due to a slow buildup of vortices \cite{bean_surface_1964, zeldov_geometrical_1994}. Superimposed on this trend, we observe noise attributed to flux avalanches \cite{nulens_catastrophic_2023}, clearly discernible in \autoref{fig:field}(c) as jumps in the \subtxt{f}{r} dependence as the field is swept from zero towards positive/negative values. When sweeping the field back towards zero, vortices are rapidly annihilated \cite{bean_surface_1964, zeldov_geometrical_1994} by antivortices formed due to the local internal field reversing polarity before the external field crosses zero \cite{nulens_catastrophic_2023}. The decreasing vortex density causes a sharp rise in resonance frequency, which reaches a maximum just before the applied field returns to zero, indicating a point of minimal vortex density.

Concurrently, the quality factor initially increases with field due to trapping of quasiparticles by vortices. However, as the vortex density grows, their dissipative effect outweighs the benefits of quasiparticle trapping, and the quality factor begins to decline. Upon sweeping the field back to zero, the quality factor drops even further, although the strong frequency increase suggests efficient vortex annihilation. We interpret this somewhat uncommon observation as an indication that vortices in our system are highly effective at suppressing quasiparticle-induced loss. Their rapid removal therefore enhances the effect of quasiparticle dissipation. The quality factor starts to rise at small fields, when the quasiparticle density decreases and their trapping becomes more efficient.

\section{Conclusion}
We demonstrate high-quality microwave resonators in a \gls{ge}-based hybrid \gls{hs}. Unlike previous implementations of such resonators, our \gls{ge}/\sige\ \gls{qw} structure is grown on a \gls{ge} substrate, and the \gls{al} film is deposited in-situ, via \gls{mbe}, as an integral part of the hybrid \gls{hs}. This approach enables to grow a compact semiconductor part of the hybrid \gls{hs} without introducing a lossy \gls{ge}-on-\gls{si} interface. The in-situ \gls{mbe} growth also facilitates the realization of ultra-thin \gls{al} films with a high kinetic inductance. While this is not the focus of this study, we suspect the in-situ integration of the hybrid \gls{hs} to additionally provide a clean superconductor–semiconductor interface (see \gls{tem} in Supplementary Information).

We extract internal quality factors, with $\subtxt{Q}{i} \approx 49000$ in the quantum regime (single-photon), $\subtxt{Q}{i} > 20000$ in the unsaturated \gls{tls} regime, and we measure \subtxt{Q}{i} up to $\subtxt{Q}{i} \approx 60000$ at high power. These values suggest long coherence times even when integrated with a quantum device. The resonators are compatible with large in-plane magnetic fields corresponding to Zeeman energies of \SI{10}{\micro\electronvolt} for typical in-plane $g$-factors in \gls{ge} quantum dots \cite{hofmann_assessing_2019, jirovec_dynamics_2022}. We also characterize the hysteresis in resonance frequency and quality factor under out-of-plane magnetic fields, indicating a complex interplay between quasiparticles and vortices.

\section{Funding statement}
This work was supported as a part of NCCR SPIN, a National Centre of Competence in Research, funded by the Swiss National Science Foundation (grant number 225153), by the European Union’s Horizon 2020 research and innovation program through the Marie Sk\l{}odowska Curie COFUND project QUSTEC, agreement No 847471, by the Swiss National Science Foundation through grant 192027 and by the Basel QCQT PhD school. The research is part of the Munich Quantum Valley, which is supported by the Bavarian state government with funds from the Hightech Agenda Bavaria.

\clearpage
\onecolumngrid   
\appendix
\section*{Supplementary Material}

\section{Superconductor/Semiconductor hybrid heterostructure epitaxy}
\begin{figure*}[ht!]
        \centering
        \includegraphics[width = 0.5\textwidth]{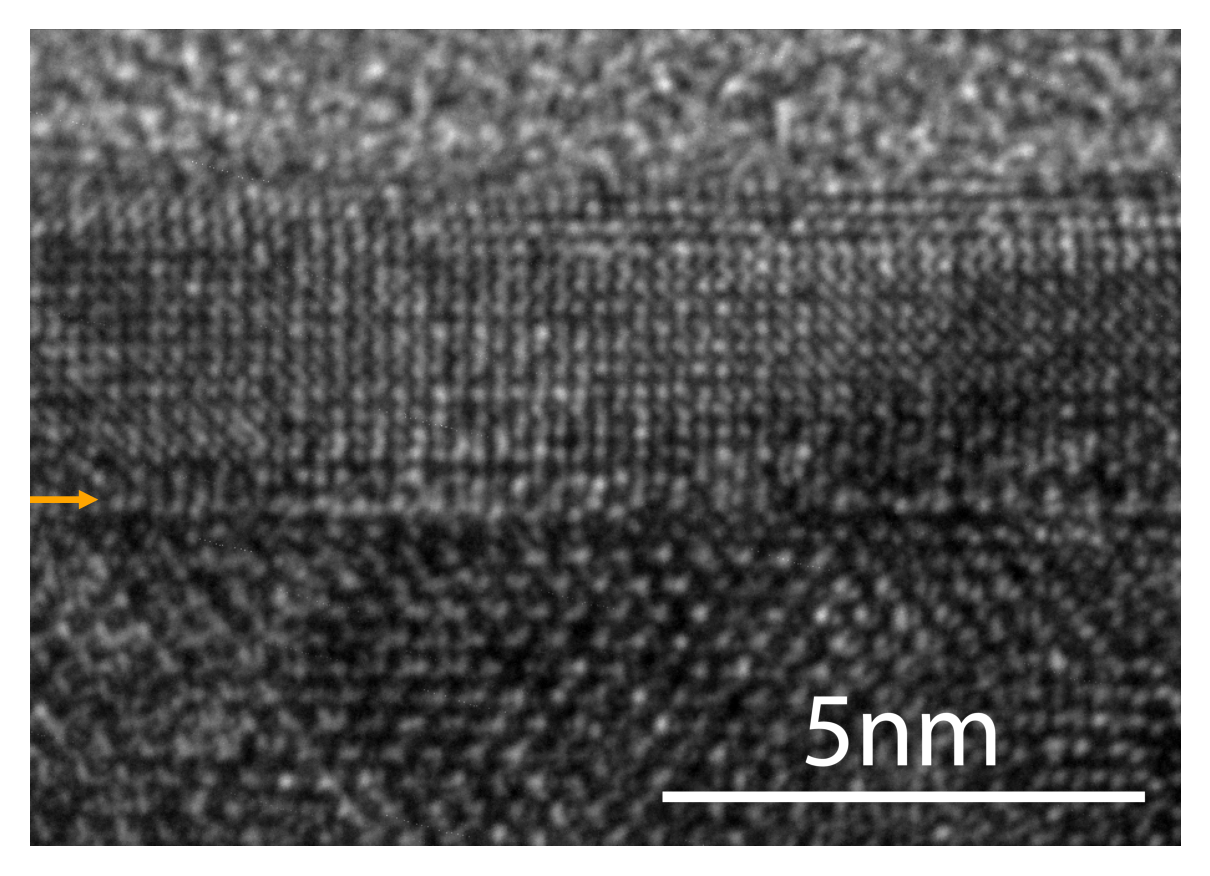}
        \caption{\textbf{TEM characterization}. HAADF image of the super/semi interface indicated by the orange arrow, with on top the \gls{al} and on the bottom the Si$_{0.05}$Ge$_{0.95}$ top barrier of the \gls{qw}.}
        \label{fig:sfigm1}
\end{figure*}

The heterostructure is grown on a 475\,\textmu m thick commercial \gls{ge}(001) substrate using solid-source \gls{mbe}. To gradually relax the strain created by lattice mismatch, a 670\,nm concentration graded Si$_{1-x}$Ge$_{x}$ buffer using the concept of linear grading of $x$ is deposited onto the substrate. The final composition is Si$_{0.2}$Ge$_{0.8}$. The graded buffer is followed by a Si$_{0.2}$Ge$_{0.8}$/Ge/Si$_{0.05}$Ge$_{0.95}$ \gls{qw}, with the respective layer thicknesses 460/15/5\,nm.\\
The superconductor/semiconductor (super/semi) hybrid heterostructure is then completed in an \gls{mbe} chamber dedicated to metal and oxide growth. The transfer between the chambers is performed without breaking the ultra-high vacuum (UHV, <10$^{-10}$\,mbar). To favor the growth of a continuous, crystalline 8\,nm \gls{al} film, the substrate is cooled to approximately $-100^{\circ}$C before the \gls{al} growth.
To prevent uncontrolled oxidation of the \gls{al} layer in air after the epitaxy, a 2\,nm aluminum oxide capping layer is deposited in situ. Subsequently, the wafer is transferred, again without breaking the UHV, into a dedicated oxidation chamber, where it is exposed to pure oxygen at $3\cdot$10$^{-6}$\,mbar for 10 minutes.\\
Fig. \ref{fig:sfigm1} shows a \gls{haadf} transmission electron micrograph (TEM) of the as-grown interface between the Si$_{0.2}$Ge$_{0.8}$/Ge/Si$_{0.05}$Ge$_{0.95}$ semiconductor \gls{qw} and the \gls{al} superconductor film.

\newpage
\section{Transmission line design}
The \gls{tl} is designed to be \SI{50}{\ohm}-matched, however considering the relation \subtxt{L}{kin} $>>$ \subtxt{L}{geo} valid for our superconducting film, this is not trivial and straightforward. A first resonator couple to a \gls{tl} was designed in order to estimate \subtxt{L}{kin}, afterwards this value was used as input in \textit{Sonnet} to estimate the optimal dimensions of the \gls{tl} in order to achieve the optimal \SI{50}{\ohm}. The value is based on what is extracted for the simulation, however, the transmission amplitude in our experiments matches well the amplification and attenuation of the lines, proving the \gls{tl} impedance is not too far off \SI{50}{\ohm}.  

\newpage
\section{Inhomogeneous Etching}
\begin{figure*}[ht!]
        \centering
        \includegraphics[width = 1.0\textwidth]{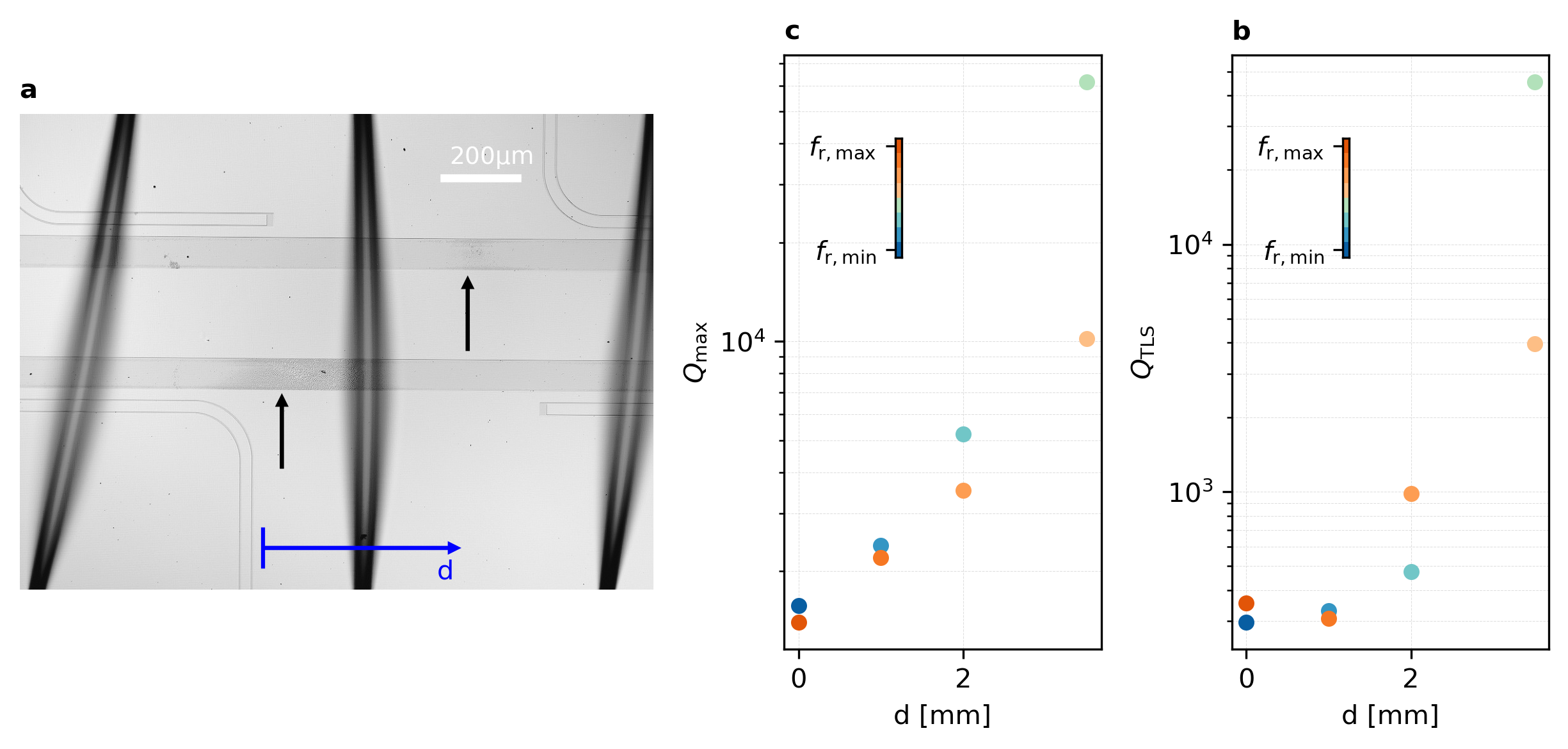}
        \caption{\textbf{Impact of inhomogeneous etching on the \subtxt{Q}{i}}. (a) Optical image of the sample with focus on the inhomogeneous etching of the \gls{tl} trenches indicated by the black arrows. (b) Max internal quality factor as a function of the distance $d$ from the "dirty spot" in (a). (c) \subtxt{Q}{TLS} defined as the difference between the \subtxt{Q}{i} at high photon number and saturated \gls{tls} and the \subtxt{Q}{i} at low photon number as a function of the distance $d$ from the "dirty spot" in (a).}
        \label{fig:sfig1}
\end{figure*}

\newpage
\section{Simulations}
\begin{figure*}[ht!]
        \centering
        \includegraphics[width = 1.0\textwidth]{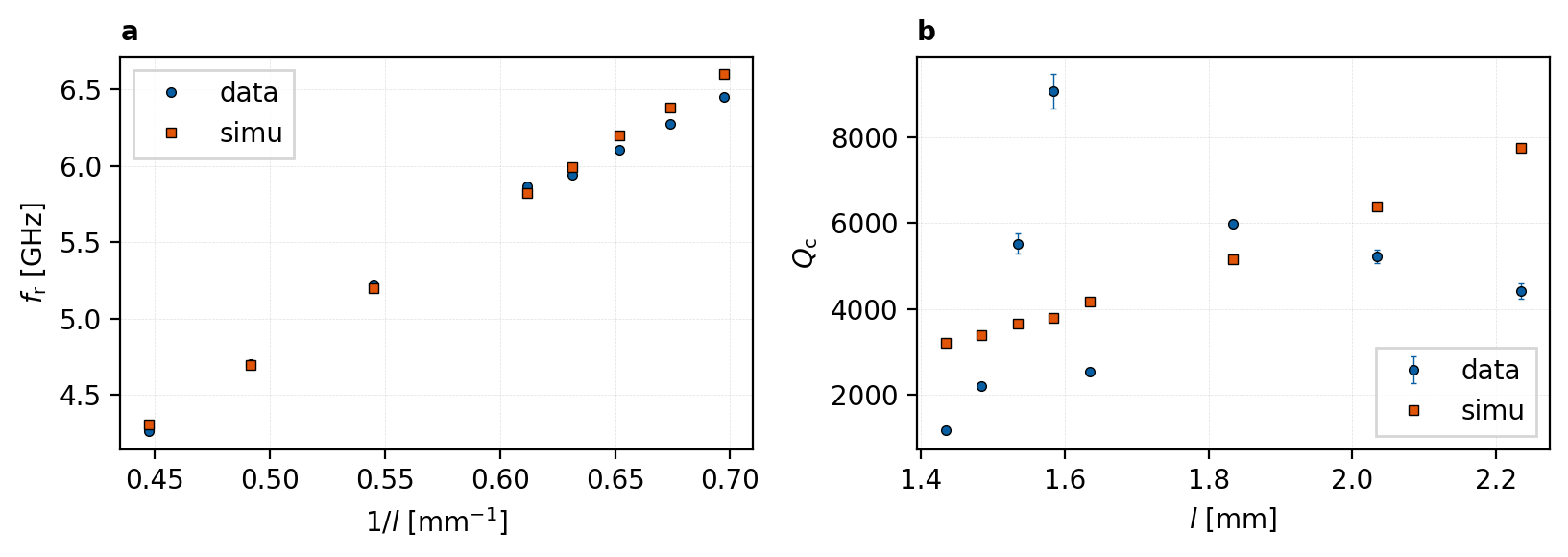}
        \caption{\textbf{Sonnet simulations}. (a) Comparison between \subtxt{f}{r} as a function of length extracted from the data and estimated from the simulations using $\subtxt{L}{kin} = \SI{36}{\pico\henry/\square}$ and \subtxt{\varepsilon}{rel} of bulk \gls{ge} of 15.8 \cite{dunlap_direct_1953} as input parameters. (b) Comparison between \subtxt{Q}{c} as a function of length extracted from the data as a function of length and estimated from the simulations. Except for some fluctuations the data are in agreement with the theory.(b) Simulated characteristics impedance \subtxt{Z}{0} as a function of frequency.} 
        \label{fig:sfig2}
\end{figure*}

%

\bibliography{resonator}

\end{document}